\documentclass{article}
\usepackage{spconf,amsmath,graphicx,mathtools,flexisym,array,float,multirow}
\newcolumntype{P}[1]{>{\centering\arraybackslash}p{#1}}
\usepackage[utf8]{inputenc}
\usepackage{comment}
\usepackage[english]{babel}
\usepackage{color}
\usepackage{framed}
\usepackage{cite}
\usepackage{amsmath,amssymb,amsfonts}
\usepackage{IEEEtrantools}
\usepackage{graphicx}
\usepackage{upgreek}
\usepackage{gensymb}
\usepackage{textcomp}
\usepackage{cite}
\usepackage{bbold}
\usepackage{ifpdf}
\usepackage{times}
\usepackage{layout}
\usepackage{float}
\usepackage{afterpage}
\usepackage{amsmath}
\usepackage{amstext}
\usepackage{amssymb,bm}
\usepackage{latexsym}
\usepackage{color}
\usepackage{kantlipsum}    
\usepackage{graphicx}
\usepackage{amsmath}
\usepackage{amsthm}
\usepackage{graphicx}
\usepackage{url}
\usepackage[font=scriptsize,caption=false,labelsep=space]{subfig}
\usepackage{booktabs}
\usepackage{multicol}
\usepackage{lipsum}

\usepackage{enumitem}
\usepackage[switch]{lineno}
\usepackage[named]{algo}
\usepackage[noend]{algpseudocode}
\usepackage{algorithm}
\usepackage{seqsplit}
\usepackage{amsmath}
\usepackage{amsthm}
\usepackage{dsfont}
\usepackage{thmtools}
\usepackage{amssymb}
\usepackage[dvipsnames]{xcolor}

\def\bPhi{\boldsymbol{\Phi}}

\def\bOmega{\boldsymbol{\Omega}}

\def\mba{\mathbf{a}}
\def\mbb{\mathbf{b}}

\def\mbn{\mathbf{n}}

\def\mbp{\mathbf{p}}

\def\mbs{\mathbf{s}}

\def\mbu{\mathbf{u}}
\def\mbv{\mathbf{v}}

\def\mby{\mathbf{y}}

\def\mbB{\mathbf{B}}
\def\mbC{\mathbf{C}}

\def\mbF{\mathbf{F}}
\def\mbG{\mathbf{G}}
\def\mbH{\mathbf{H}}
\def\mbI{\mathbf{I}}

\def\mbP{\mathbf{P}}
\def\mbQ{\mathbf{Q}}
\def\mbR{\mathbf{R}}

\def\mbT{\mathbf{T}}

\def\mbZ{\mathbf{Z}}

\def\bzero{\boldsymbol{0}}

\newcommand{\complexC}[1]{\mathds{C}^{#1}}
\newcommand{\expecE}[1]{\mathds{E}\left\{{#1}\right\}}

\newcommand{\realR}[1]{\mathds{R}^{#1}}

\def\Tr#1{\mathrm{Tr}\left(#1\right)}
\def\vec#1{\mathrm{vec}\left(#1\right)}

\def\Diag#1{\mathrm{Diag}\left(#1\right)}

\newtheorem{proposition}{Proposition}

\def\T{\top}
\def\H{\mathrm{H}}
\def\j{\mathrm{j}}

\def\baleph{\boldsymbol{\aleph}}

\newcommand*{\rom}[1]{\expandafter\@slowromancap\romannumeral #1@}

\usepackage{tikz}
\title{Quantized Phase-Shift Design of Active IRS for\\ Integrated Sensing and Communications}
\name{Zahra Esmaeilbeig$^1$$^\star$,  Arian Eamaz$^1$$^\star$, Kumar Vijay Mishra$^{2}$, and Mojtaba Soltanalian$^1$ \thanks{$^{\star}$Equal contribution. This work was sponsored in part by the National Science Foundation Grant  ECCS-1809225, and in part by the Army Research Office, accomplished under Grant Number W911NF-22-1-0263. The views and conclusions contained in this document are those of the authors and should not be interpreted as representing the official policies, either expressed or implied, of the Army Research Office or the U.S. Government. The U.S. Government is authorized to reproduce and distribute reprints for
Government purposes notwithstanding any copyright notation herein.}}
\address{$^{1}$ECE Department, University of Illinois Chicago, Chicago, IL 60607, USA\\
 $^{2}$United States DEVCOM Army Research Laboratory, Adelphi, MD 20783, USA\vspace{-16pt}}
\ninept
\begin{document}
\setlength{\abovedisplayskip}{3pt}
\setlength{\belowdisplayskip}{3pt}
\maketitle
\begin{abstract}
Integrated sensing and communications (ISAC) is a spectrum-sharing paradigm that allows different users to jointly utilize and access the crowded electromagnetic spectrum. 
In this context, intelligent reflecting surfaces (IRSs) have lately emerged as an enabler for non-line-of-sight (NLoS) ISAC. Prior IRS-aided ISAC studies assume passive surfaces and rely on the continuous-valued phase-shift model. In practice, the phase-shifts are quantized. Moreover, recent research has shown substantial performance benefits with active IRS. In this paper, we include these characteristics in our IRS-aided ISAC model to maximize the receive radar and communications signal-to-noise ratios (SNR) subjected to a unimodular IRS phase-shift vector and power budget. The resulting optimization is a highly non-convex unimodular quartic optimization problem. We tackle this problem via a bi-quadratic transformation to split the design  into two quadratic sub-problems that are solved using the \emph{power iteration} method. The proposed approach employs the $M$-\emph{a}ry unimodular sequence design via \emph{r}elaxed power method-\emph{li}ke iteration (MaRLI) to design the quantized phase-shifts. Numerical experiments employ continuous-valued phase shifts as a benchmark and demonstrate that our active-IRS-aided ISAC design with MaRLI converges to a higher value of SNR with an increase in the number of IRS quantization bits. 
\end{abstract}
\vspace{-6pt}
\vspace{-8pt}
\section{Introduction}
\vspace{-8pt}
Next-generation communications are expected to provide significant performance enhancements to meet the demands of emerging applications such as vehicular networks, smart warehouses, and virtual/augmented reality with high throughput and low latency \cite{boccardi2014five}. This requires a judicious sharing of the electromagnetic spectrum, which is a scarce resource, by both incumbent and opportunistic users \cite{mishra2019toward}. In this context, integrated sensing and communications (ISAC) offers significant advantages over traditional wireless systems by combining sensing and communications functions into a single device and a joint waveform to prevent mutual interference \cite{hassanien2019dual}.

A recent trend in ISAC research is employing intelligent reflecting surfaces (IRSs) to enable non-line-of-sight (NLoS) sensing and communications \cite{elbir2022rise,wei2022multi,wei2022multiple,wei2022simultaneous,wang2023stars}. An IRS comprises several subwavelength units or meta-atoms that are able to manipulate incoming electromagnetic waves in a precisely controlled manner through modified boundary conditions \cite{hodge2020intelligent,hodge2023index}. Recent studies have shown that IRS exploits NLoS signals to extend the coverage area and bypass line-of-sight (LoS) blockages in radar \cite{esmaeilbeig2022cramer,esmaeilbeig2022joint,esmaeilbeig2022irs} and communications \cite{zheng2022survey}. 

Prior IRS-aided ISAC research assumes IRS as a passive device, whose continuous-valued phase-shifts need to be optimized~\cite{wang2023stars,wei2022multiple,mishra2022optm3sec}. However, in hardware implementations, IRS phases are quantized \cite{di2020hybrid,you2020intelligent,gao2021robust}. While there is some literature \cite{wei2022irs,wang2021joint} on using quantized IRS for ISAC, they model IRS as a completely passive device. In general, IRS may be equipped with active RF components that allow changing the amplitude of the incoming signal, among other functionalities \cite{hodge2021deep,mishra2023machine}. 
An active IRS consumes less power and has low processing latency in comparison to a relay~\cite{xu2021resource,mishra2023machine,zhang2023active}. In this paper, we impose constraints on both the transmitter and IRS power leading to a design criterion that allocates only a fraction of the transmit power to the IRS. Contrary to previous research, we employ active IRS with quantized phase-shifts for the ISAC system. In particular, we focus on optimizing the DFBS precoder and active IRS parameters. In the case of continuous-valued passive IRS-ISAC \cite{wei2022irs,li2022dual}, this joint optimization problem is highly non-convex because of the unit modulus constraint or \emph{unimodularity} on each element of the IRS parameter matrix. In such cases, the problem is cast as a unimodular quadratic program (UQP)~\cite{soltanalian2014designing} which is NP-hard when the phase-shifts are quantized. A semi-definite program (SDP) may relax the problem but it is computationally expensive \cite{li2023minorization,eamaz2022phase}. Recently, power method-like iteration (PMLI) algorithms, inspired by the power iteration method's advantage of simple matrix-vector multiplications \cite{soltanalian2014designing,eamaz2023cypmli}, have been shown to address 
 UQPs efficiently~\cite{soltanalian2014designing}. 

We cast the IRS-ISAC quantized phase-shifts design as a unimodular quartic program (UQ$^2$P) that we split into two low-complexity quadratic sub-problems through a quartic to bi-quadratic transformation~\cite{eamaz2023marli,esmaeilbeig2022joint,eamaz2023near}. We then tackle each quadratic sub-problem with respect to the quantized phase-shifts using the recently proposed $M$-ary unimodular sequence design via relaxed PMLI (MaRLI) algorithm \cite{eamaz2023marli}. Here, the conventional projection operator of the exponential function in the PMLI is replaced by a \emph{relaxation operator} \cite{soltanalian2012computational,eamaz2023marli}. This ensures enhanced convergence to the desired discrete set. Our numerical experiments show that the proposed algorithm increases the signal-to-noise ratio (SNR) even with quantized phase-shifts.  

Throughout this paper, we use bold lowercase and bold uppercase letters for vectors and matrices, respectively. The $mn$-th element of the matrix $\mbB$ is $\left[\mbB\right]_{mn}$. $(\cdot)^{\top}$, $(\cdot)^{\ast}$and $(\cdot)^{\mathrm{H}}$ are the vector/matrix transpose, conjugate and the Hermitian transpose, respectively; The trace of a matrix is denoted by$\operatorname{Tr}(.)$; the function $\textrm{diag}(.)$ returns the diagonal elements of the input matrix, while $\textrm{Diag}(.)$ produces a diagonal matrix with the same diagonal entries as its vector argument. The  Kronecker and Hadamard products are denoted by $\otimes$ and $\odot$, respectively. The vectorized form of a matrix $\mbB$ is written as $\vec{\mbB}$. The $s$-dimensional  all-zeros vector, and the identity matrix of  size $s\times s$ are  $\mathbf{0}_{N}$, and $\mbI_s$, respectively.  For any real number $x$, the function $[x]$ yields the closest integer to $x$ (the largest is chosen when this integer is not unique) and $\{x\}=x-[x]$. 
\vspace{-6pt}
\section{Signal model}
\begin{figure}[t]
\centering
	\input{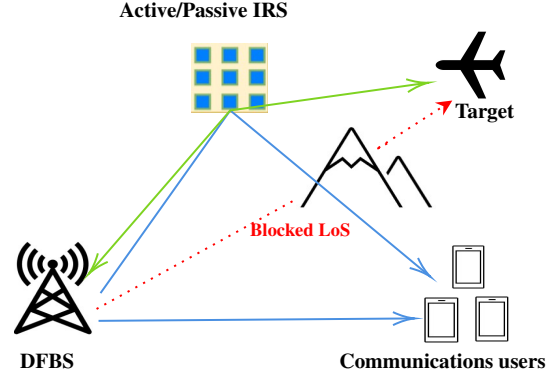}
     \caption{\footnotesize{A simplified illustration of IRS-aided ISAC system. When the LoS is blocked, the NLoS paths via the IRS allow for establishing the link between the targets/users with the DFBS. \vspace{-16pt}}}  
\label{fig_1}
\end{figure}

\vspace{-8pt}
Consider an IRS-aided ISAC system (Fig.~\ref{fig_1}) that consists of a dual-function base station (DFBS) with $N$ elements each in transmit (Tx)  and receive (Rx) antenna arrays. The IRS comprises $L=L_x \times L_y$  reflecting elements arranged as a uniform planar array (UPA) with $L_x$ ($L_y$) elements along the x- (y-) axes in the Cartesian coordinate plane. Define the IRS steering vector 
$\mba(\theta_h,\theta_v)=\mba_x(\theta_h,\theta_v) \otimes \mba_x(\theta_h,\theta_v)$, where $\theta_h$ ($\theta_v$) is the azimuth (elevation) angle, $\mba_x(\theta_h,\theta_v)=[1,e^{\textrm{j}\frac{2\pi d}{\lambda} cos \theta_h sin \theta_v},\ldots,e^{\textrm{j}\frac{2\pi d (L_x-1)}{\lambda} cos \theta_h sin \theta_v}]^{\top}$ and  $\mba_y(\theta_h,\theta_v)=[1,e^{\textrm{j}\frac{2\pi d}{\lambda} cos \theta_h sin \theta_v},\ldots,e^{\textrm{j}\frac{2\pi d (L_y-1)}{\lambda} cos \theta_h sin \theta_v}]^{\top}$, $\lambda=c/f_c$ is the carrier wavelength, $c=3 \times 10^8$ m/s is the speed of light, $f_c$ is the carrier frequency, and $d=0.5\lambda$ is the  inter-element (Nyquist) spacing. The IRS operation is characterized by the parameter matrix
$\bPhi=\Diag{\mbv}=\Diag{[b_1e^{\j\phi_1},\ldots,b_Le^{\j\phi_L}]}=\Diag{\mbb \odot \mbu}$, where  $\mbb=[b_1,\ldots,b_L]^{\T}$ and  $\mbu=[e^{\j\phi_1},\ldots,e^{\j\phi_L}]$ are gain and phase-shift vectors, respectively. 

For active (passive) IRS we have $|b_l|>1$ ($|b_l|=1$), i.e., a passive IRS modifies only the phase shift of the impinging waveform. For IRS phase-shifts with with $M$ quantization-levels, the feasible set of $\mbu$ is the set of polyphase sequences 
\par\noindent\small
\begin{align}\label{ST1}
\Omega^{L}_{M}&= \left\{\mbu \in \complexC{L}| u_l= e^{\textrm{j}\omega_{l}}, \omega_l \in \Psi_{M},~0\leq l\leq L-1\right\},
\end{align} \normalsize
where $\Psi_{M}=\left\{1,\frac{2\pi}{M},\cdots,\frac{2\pi(M-1)}{M}\right\}$ is quantized phase-shift set. 

Assume that the DFBS transmits an orthogonal symbol vector  $\mbs=[s_1,\ldots,s_K]^{\T}$ where $\expecE{\mbs\mbs^{H} } =\mbI_{K}$ to $K$ communications users and sense a target. The Tx and active IRS powers  are $P_{_T}$ and $P_{_{\textrm{IRS}}}$, respectively. The continuous-time  transmit signal from $n_t$-th Tx antenna is 
\begin{equation}
x_{_{n_t}}(t)= \sum_{k=1}^{K}[\mbP]_{_{n_t,k}} s_{_{k}} \textrm{rect}(t-k\Delta t)e^{\j2\pi f_c t},\; 0 <t<K\Delta t,
\end{equation}
where $[\mbP]_{_{n_t,k}}$ is the $(n_t,k)$-th element of the DFBS precoder 
$\mbP\in \complexC{N \times K}$, $\Delta t$ is the symbol duration,  
and
$\textrm{rect}(t)=\begin{cases}
  1 & \text{if} \; |t|<\frac{1}{2}, \\
  0 & \text{otherwise}.
\end{cases}$. The covariance of the DFBS Tx signal is $\mbR_D=
\mbP\mbP^{\H}$.

\noindent\textbf{Communications Rx signal:} In communications  setup, denote the direct  channel state information (CSI) and IRS-reflected non-line-of-sight (NLoS) CSI matrices by $\mbF \in \complexC{N \times N}$ and $\mbH \in \complexC{K \times L}$, respectively. Then, at each communications receiver after  sampling, the discrete-time received signal is $\mby_{c,k}$. Concatenating the signal of all  users, we obtain the  $K \times 1$ vector: 
\begin{equation}
\mby_c=(\mbF+\mbH\bPhi\mbG) \mbP \mbs +  \mbn_c,
\end{equation}
where the  Tx-IRS CSI matrix $\mbG \in \complexC{L \times N}$ is assumed to be estimated \textit{a priori} through suitable channel estimation techniques \cite{wei2022multi} and  $\mbn_c \sim \mathcal{N}(\bzero,\sigma^2_{c}\mbI_{K})$  is  the noise  at communications receivers. 

\noindent\textbf{DFBS Rx signal:} Consider the NLoS Swerling-0 \cite{skolnik2008radar} radar target located at range $r_t$ with  respect to the  DFBS and 
direction-of-arrival (DoA) $(\theta_{h_t},\theta_{v_t})$ with respect to the IRS and radar cross-section (RCS) $\alpha_{_T}$. Define $\mbR=\alpha_{_T} \mbG^{T}\bPhi \mba(\theta_{h_t},\theta_{v_t}) \mba(\theta_{h_t},\theta_{v_t})^{\T}\bPhi \mbG$. The continuous-time baseband signal at $n_r$-th DFBS Rx antenna is 
\par\noindent\footnotesize
\begin{equation}
y_{_{n_r}}(t)=
\sum_{n_t=1}^{N} [\mbR]_{_{n_r,n_t}}x_{_{n_t}}(t-\tau),\; 0<t<K\Delta t,
\end{equation}\normalsize
where $\tau=\frac{2r_t}{c}$ is the range-time delay and $[\mbR]_{_{n_r,n_t}}$ 
accounts for  the RCS and DoA information of the target with respect to the $n_t$-th transmit and $n_r$-th receive antenna.  Stacking the echoes for all receiver antennas, the $N \times 1$  
signal at radar receiver after downconversion and  sampling is $\mby_{_r}=[y_{_{1}}(t), \ldots, y_{_{N}}(t)]^{\T}
= \mbR\mbP \mbs +\mbn_r$, where $\mbn_r \sim \mathcal{N}(\bzero,\sigma^2_{r}\mbI_{N})$  is  the noise  at radar receiver.
Denote  $\mbC=\mbF+\mbH\bPhi\mbG$. The output SNR at communications and DFBS Rx 
are, respectively, 
\par\noindent\footnotesize
 \begin{align}\label{snr_c}
 \textrm{SNR}_{c}&=\frac{1}{\sigma^2_{c}}\Tr{\mbC \mbP \mbP^{\H}\mbC^{\H}},
 \end{align}\normalsize
and\par\noindent\footnotesize
 \begin{align} \label{snr_r}
 \textrm{SNR}_{r}&=\frac{1}{\sigma^2_{r}}\Tr{\mbR \mbP \mbP^{\H}\mbR^{\H}}.
\end{align}\normalsize
The following Proposition states $\textrm{SNR}_{r}$ as a quartic function with  respect to the IRS phase shift vector $\mbv$ or equivalently with respect to  $\mbb$ and $\mbu$. It further expresses 
$\textrm{SNR}_{c}$ as a  quadratic function with  respect to  IRS complex gain.  
Hereafter, we assume the  DFBS is scanning a specified  azimuth-elevation bin of  the environment for the target. The DoA $(\theta_h,\theta_v)$ is estimated by solving a separate optimization problem that we omit in this work because of the paucity of space. 
and  denote $\mba(\theta_{h},\theta_{v})$ by  $\mba$, for  brevity. 
 \begin{proposition}
 \label{St_Pro}
 Define the variables $\tilde{\mbF}=\vec{\mbF}$, 
 $\tilde{\mbG}=\mbG^{\T}\otimes \mbH$, 
 $\hat{\mbQ}=(\mbP^{\T}\otimes \mbI_{K})^{\H}(\mbP^{\T}\otimes \mbI_{K})$,
$T=[\vec{\baleph_1}~\vdots~\ldots~\vdots~\vec{\baleph_L}]$, and $\baleph_{l}$ as an $L \times L$ matrix with $[\baleph]_{_{ll}}=1$ and zero everywhere else. Denote 
$\boldsymbol{\upalpha}=\mbT^{\H}\tilde{\mbG}^{\H}\hat{\mbQ}\tilde{\mbF}$ and $\tilde{\mbQ}=\mbT^{\H}\tilde{\mbG}^{\H}\hat{\mbQ}\tilde{\mbG}\mbT$. Then, 
\begin{equation}
\label{St10}
\text{SNR}_{c}=\bar{\mbv}^{\H}  \tilde{\bOmega} \bar{\mbv},
\end{equation}
where $\bar{\mbv}= \bar{\mbb} \odot \bar{\mbu}$, with $ \bar{\mbb}=  [\mbb^{\T} 1]^{\T}$, $ \bar{\mbu}=  [\mbu^{\T} 1]^{\T}$, and  $\tilde{\bOmega}=\frac{1}{\sigma^2_{c}}\begin{bsmallmatrix} \tilde{\mbQ} & \boldsymbol{\upalpha} \\ \boldsymbol{\upalpha}^{\H} & \tilde{\mbF} \hat{\mbQ} \tilde{\mbF}
\end{bsmallmatrix}$. 
Then, 
\begin{align}\label{St11}
\text{SNR}_{c} = \bar{\mbu}^{\H}\tilde{\bOmega}^{(1)}\bar{\mbu}
= \bar{\mbb}^{\H}\tilde{\bOmega}^{(2)}\bar{\mbb},
\end{align}
where $\tilde{\bOmega}^{(1)}=\bar\mbb^{*}\bar\mbb^{\T} \odot \tilde{\bOmega}$ and $\tilde{\bOmega}^{(2)}=\bar\mbu^{*}\bar\mbu^{\T} \odot \tilde{\bOmega}$.
Similarly, 
\begin{equation}
\label{St8}
\textrm{SNR}_{r}=\mbv^{H}\mbQ(\mbv)\mbv,
\end{equation}
where $\mbv=\mbb \odot \mbu$ , $\mbQ(\mbv)=\frac{|\alpha_{_T}|^{2}}{\sigma^2_{r}}\left (\mbv^{\H}\bOmega^{\star}\otimes\bOmega^{\star}\mbP^{\star}\right)\left(\bOmega^{\T}\mbv \otimes \mbP^{\T}\bOmega^{\T}\right)$ and $\bOmega=\Diag{\mba}\mbG$. Then,  \eqref{St8} is rewritten as 
\begin{equation}
 \label{St9}
 \begin{aligned}
  \textrm{SNR}_{r} = \mbu^{\H}\mbQ^{(1)}(\mbv)\mbu   = \mbb^{\H}\mbQ^{(2)}(\mbv)\mbb,
 \end{aligned}
 \end{equation}
 where $\mbQ^{(1)}(\mbv)=\mbb^{*}\mbb^{\T} \odot \mbQ(\mbv)$ and $\mbQ^{(2)}(\mbv)=\mbu^{*}\mbu^{\T} \odot \mbQ(\mbv)$.
 \end{proposition}
 \begin{IEEEproof} Consider the following expression in $\textrm{SNR}_c$: 
 \par\noindent\footnotesize
 \begin{align}\label{eq:tran}
 &\Tr{\mbC\mbP\mbP^{H}\mbC^{\H}}=\vec{\mbC \mbP}^{\H}\vec{\mbC \mbP}\nonumber\\
 &=\left ((\mbP^{\T}\otimes \mbI_{K})\vec{\mbC}\right)^{\H} (\mbP^{\T}\otimes \mbI_{K}) \vec{\mbC} \nonumber\\
 &=\mbv^{\H}\Tilde{\mbQ}\mbv + \boldsymbol{\upalpha}^{\H}\mbv+\mbv^{H}\boldsymbol{\upalpha}+ \tilde{\mbF} \hat{\mbQ} \tilde{\mbF}=\begin{bsmallmatrix} \mbv \\ 1\end{bsmallmatrix}^{\H}\begin{bsmallmatrix} \tilde{\mbQ} & \boldsymbol{\upalpha} \\ \boldsymbol{\upalpha}^{\H} & \tilde{\mbF} \hat{\mbQ} \tilde{\mbF}
\end{bsmallmatrix}\begin{bsmallmatrix} \mbv \\ 1\end{bsmallmatrix}
 \end{align}\normalsize
 where we used  
$ \vec{\bPhi}= \vec{\Diag{\mbv}}=\mbT \mbv$ \cite[Lemma 1]{esmaeilbeig2023moving}  and 
$\vec{\mbC}= \vec{\mbF}+(\mbG^{\T}\otimes \mbH)\vec{\bPhi}=\tilde{\mbF}+\tilde{\mbG}\mbT\mbv$.
Substituting $\mbv=\Diag{\mbb}\mbu=\Diag{\mbu}\mbb$ in~\eqref{St10} will  result in~\eqref{St11}. 
The proofs of \eqref{St8} and \eqref{St9} follow, \textit{mutatis mutandis}, through \eqref{snr_r}. 
 \end{IEEEproof}
Our goal is  to jointly design the  IRS and  DFBS precoder  matrix to maximize the SNR at communications and DFBS Rx. 
\section{Problem formulation}
\vspace{-8pt}
Prior studies on IRS-ISAC design are either radar- or communications-centric choosing to optimize either of the systems while constraining the performance of the other. For instance, \cite{wang2023stars} minimizes the trace of the radar target parameter Cram\'{e}r–Rao lower bound matrix subject to a minimum communications user SNR. Here, we adopt an equitable approach to optimize the SNRs of both systems.  


Define the  weighted sum of the radar and communications SNRs, with a weight factor $\beta$ as $\textrm{SNR}_{_T}=\beta \textrm{SNR}_{r} + (1-\beta) \textrm{SNR}_{c} $  as the design criterion and the power allocated to Tx and  IRS as constraints. 
Our  design problem is \par\noindent \small
\begin{align}\label{eq:opt1}
\mathcal{P}_{1}:&\;\underset{\mbu,\mbb, \mbP}{\textrm{maximize}} \; \textrm{SNR}_{_T} \nonumber \\
&\;\textrm{subject to} \;
 \mbP \mbP^{\H}=\mbR_{D},\;
   \left\|\mbP\right\|^{2}_{\mathrm{F}}&= P_{_T}, \; \left\|\mbb\right\|^{2}_{2}=P_{_{\textrm{IRS}}},
\end{align}\normalsize
The problem $\mathcal{P}_{1}$ is  highly nonconvex because of the coupling between quantized phase-shifts and precoder parameters. 
We, therefore, solve it via a cyclic optimization over each design parameter as detailed below. 
 
\noindent\textbf{IRS Design:} Using ~\eqref{St10} and ~\eqref{St8} from Proposition~\ref{St_Pro}, the IRS beamforming design problem is formulated as \par\noindent\small
\begin{equation}\label{eq:bar_v}
\mathcal{P}_{_2}:~\underset{\bar \mbv}{\textrm{maximize}}\quad \bar{\mbv}^{\H}  \check{\bOmega} \bar{\mbv},
\end{equation}\normalsize
where $\check{\bOmega}=\beta \begin{bsmallmatrix} \mbQ(\mbv) & \bzero_{_L} \\ 0 &0  
\end{bsmallmatrix}+(1-\beta)\tilde{\bOmega}$ 
and $\bar{\mbv}= \bar{\mbb} \odot \bar{\mbu}$  is comprised of the vector of  amplitudes $\mbb$ and quantized phase-shifts $\mbu$. Consequently, \eqref{eq:bar_v} is split into two quartic sub-problems that are cyclically 
solved with respect to $\mbb$ and $\mbu$. From~\eqref{St11}, $\mathcal{P}_{_2}$ with respect to $\mbb$ is 
\par\noindent \small
\begin{align}\label{eq:opt}
\mathcal{P}^{(1)}_{_2}:~\underset{\mbb}{\textrm{maximize}}\;  \bar{\mbb}^{\H}  \check{\bOmega}^{(1)} \bar{\mbb} \quad
\textrm{subject to} \;  \left\|\mbb\right\|^{2}_{2}=P_{_{\textrm{IRS}}},
\end{align}\normalsize
where $\check{\bOmega}^{(1)}=\bar\mbu^{*}\bar\mbu^{\T} \odot \check{\bOmega}$. Further, $\mathcal{P}_{_2}$ with respect to $\mbu$ is
\begin{equation}
\label{St12}
\begin{aligned}
\mathcal{P}^{(2)}_{_2}:~\underset{\mbu\in \Omega^{L}_{M}}{\textrm{maximize}}& \quad \bar{\mbu}^{\H}  \check{\bOmega}^{(2)} \bar{\mbu},
\end{aligned}
\end{equation}
where $\check{\bOmega}^{(2)}= \bar\mbb^{*}\bar\mbb^{\T} \odot \check{\bOmega}$. Passive IRS does not require $\mathcal{P}^{(1)}_{2}$. 

\noindent\textbf{DFBS Precoder Design:} 
Substituting \eqref{snr_r} and \eqref{snr_c} into \eqref{eq:opt1}, problem  $\mathcal{P}_{1}$ with respect to  $\mbP$ becomes equivalent to \par\noindent\footnotesize
\begin{equation}
\begin{aligned}
\label{St3}
\mathcal{P}_{3}:~\underset{\mbP}{\textrm{maximize}}& \quad \Tr{\mbP\mbP^{\H}\mbZ} \\
\textrm{subject to}
& \quad\left\|\mbP\mbP^{\H}-\mbR_D\right\|^2_{\mathrm{F}}\leq \eta,~ \left\|\mbP\right\|^{2}_{\mathrm{F}} = P_{_T}
\end{aligned}
\end{equation}\normalsize
where $\mbZ=\frac{\beta}{\sigma^2_r}\mbR^{\H}\mbR+\frac{1-\beta}{\sigma^2_c}\mbC^{\H}\mbC$ and $\eta$ is a positive constant. Define $\check \mbZ=(\mbI_K \otimes \mbZ)$ and $\tilde{\mbp}=\vec{\mbP}$. If one employs same algebraic transformations as in~\eqref{eq:tran}, we have $\Tr{\mbP\mbP^{\H}\mbZ}
= \tilde{\mbp}^{\H}\check \mbZ\tilde{\mbp}$.

Assume $\lambda_{m}$ is the maximum eigenvalue of $\check \mbZ$, where $\lambda_{m}\mbI\succeq \check \mbZ$. We   deploy \emph{diagonal loading} to  replace $\check \mbZ$ with $\breve{\mbZ}$. One can verify that \emph{diagonal loading} with  $\breve{\mbZ} = \lambda_{m}\mbI - \check \mbZ$  will not change the solution and only changes the maximization to minimization and  ensures that  $\breve{\mbZ}$ is positive semidefinite.
\par\noindent\footnotesize	
\begin{align}
\label{St4}
\mathcal{P}^{(1)}_{3}:~\underset{\mbP}{\textrm{minimize}}& \quad \tilde{\mbp}^{\H}\breve{\mbZ}\tilde{\mbp} +\gamma\left\|\mbP\mbP^{\H}-\mbR_D\right\|^2_{\mathrm{F}} \\
\textrm{subject to}
&\quad \left\|\mbP\right\|^{2}_{\mathrm{F}} =\left\|\tilde{\mbp}\right\|^{2}_{2}= P_{_T},
\end{align}\normalsize
where $\gamma$ is the Lagrangian multiplier. Reformulate 
$\left\|\mbP\mbP^{\H}-\mbR_D\right\|^2_{\mathrm{F}}$ as $\tilde{\mbp}^{\H}\left(\mbI_K \otimes \vec{\mbP\mbP^{\H}}\right)\tilde{\mbp}
-2\tilde{\mbp}^{\H}(\mbI_K \otimes \mbR_{D})\tilde{\mbp}+\mbR^{\H}_D\mbR_{D}$. Consequently, we obtain the following quartic program
\begin{align}
\label{St6}
\mathcal{P}^{(1)}_{3}:~\underset{\tilde{\mbp}}{\textrm{minimize}}\;
\tilde{\mbp}^{\H}\breve{\bOmega}\left(\tilde{\mbp}\right)\tilde{\mbp}
\quad
\textrm{subject to}
\; \left\|\tilde{\mbp}\right\|^{2}_{2}= P_{_T},
\end{align}
where  $\breve{\bOmega}\left(\tilde{\mbp}\right)=\breve{\mbZ}+\gamma\left(\mbI_K \otimes \vec{\mbP\mbP^{\H}}\right)-2\gamma (\mbI_K \otimes \mbR_{D})$. Using the diagonal loading, we change \eqref{St6} to a maximization problem. Assume $\lambda_{M}$ is the maximum eigenvalue of $\breve{\bOmega}$.
Thus, $\hat{\bOmega} = \lambda_{M}\mbI -\breve{\bOmega}$ is positive definite and we get
\begin{align}
\label{St7}
\mathcal{P}^{(2)}_{3}:~\underset{\tilde{\mbp}}{\textrm{maximize}}\;
\tilde{\mbp}^{\H} \hat{\bOmega}\left(\tilde{\mbp}\right) \tilde{\mbp} 
\quad
\textrm{subject to}\;
 \left\|\tilde{\mbp}\right\|^{2}_{2}= P_{_T},
\end{align}
Note that a diagonal loading with $\lambda_{M}\mbI$ has no effect on the solution of (\ref{St7}) because $\tilde{\mbp}^{\mathrm{H}} \hat{\bOmega}\left(\tilde{\mbp}\right) \tilde{\mbp}=\lambda_{M}P_{0}-\tilde{\mbp}^{\mathrm{H}} \breve{\bOmega}\left(\tilde{\mbp}\right) \tilde{\mbp}$.

\section{Proposed Algorithm}
\vspace{-8pt}
In this section, we employ diagonal loading to turn each transform into a quadratic form suitable to apply power iteration methods.

\noindent\textbf{To tackle $\mathcal{P}^{(1)}_{_2}$ for $\mbb$ and $\mbu$ cyclically:} We resort to a task-specific \emph{alternating optimization} (AO) or \emph{cyclic optimization algorithm} \cite{bezdek2003convergence,tang2020polyphase,soltanalian2013joint}. We split $\mathcal{P}^{(1)}_{_2}$ into two quadratic optimization sub-problems with respect to $\mbb$.
Define variables $\bar{\mbb}_{1}= [\mbb^{\T}_{1} ~1]^{\T}$ and $\bar{\mbb}_{2}= [\mbb^{\T}_{2} ~1]^{\T}$. This changes $\mathcal{P}^{(1)}_{2}$ to 
\par\noindent\footnotesize
\begin{align}\label{St16}
\mathcal{P}^{(3)}_{_2}:~\underset{\mbb_{j}}{\textrm{maximize}} & \quad \bar{\mbb}^{\H}_{j}  \check{\bOmega}^{(1)}(\mbb_{i}) \bar{\mbb}_{j},\quad i\neq j \in \left\{1,2\right\}, \\
\textrm{subject to} & \quad \left\|\mbb_{1}\right\|^{2}_{2}=\left\|\mbb_{2}\right\|^{2}_{2}=P_{_{\textrm{IRS}}},
\end{align}\normalsize
If either $\mbb_1$ or $\mbb_2$ is fixed, minimizing the objective with respect to the other variable is achieved via quadratic programming. To guarantee that $\mathcal{P}^{(3)}_{_2}$ leads to $\mathcal{P}^{(1)}_{_2}$, we must show that $\mbb_1$ and $\mbb_2$ are convergent to the same value. Adding the second norm error between $\mbb_{1}$ and $\mbb_{2}$ as a \emph{penalty} with the Lagrangian multiplier to $\mathcal{P}^{(3)}_{_2}$ results in the following \emph{regularized Lagrangian} problem\cite{eamaz2023cypmli,eamaz2023near}:
\par\noindent\small
\begin{align}\label{St17}
\mathcal{P}^{(4)}_{_2}:~\underset{\mbb_{j}}{\textrm{minimize}} & \quad \bar{\mbb}^{\H}_{j}  \left(\lambda^{\prime}\mbI-\check{\bOmega}^{(1)}(\mbb_{i})\right) \bar{\mbb}_{j}+\uptau \left\|\bar{\mbb}_{i}-\bar{\mbb}_{j}\right\|^{2}_{2}, \nonumber \\
\textrm{subject to} & \quad \left\|\mbb_{1}\right\|^{2}_{2}=\left\|\mbb_{2}\right\|^{2}_{2}=P_{_{\textrm{IRS}}},
\end{align}\normalsize
where $\lambda^{\prime}$ is the maximum eigenvalue of $\check{\bOmega}^{(1)}$, and $\uptau$ is the Lagrangian multiplier. Also, $\mathcal{P}^{(4)}_{_2} $ is recast as\par\noindent\small
\begin{align}\label{St18}
\mathcal{P}^{(4)}_{_2}:\quad\underset{\mbb_{j}}{\textrm{minimize}}&\quad
\begin{bsmallmatrix} \bar{\mbb}_{j}\\1\end{bsmallmatrix}^{\mathrm{H}}
\underbrace{\begin{bsmallmatrix} \lambda^{\prime}\mbI-\check{\bOmega}^{(1)}(\mbb_{i}) &-\uptau\bar{\mbb}_{i} \\ -\uptau \bar{\mbb}_{i}^{\mathrm{H}} & 2\uptau P_{_{\textrm{IRS}}} \end{bsmallmatrix}}_{\mathcal{E}(\mbb_{i})}
\begin{bsmallmatrix} \bar{\mbb}_{j}\\1\end{bsmallmatrix},\nonumber\\
\textrm{subject to} & \quad \left\|\mbb_{1}\right\|^{2}_{2}=\left\|\mbb_{2}\right\|^{2}_{2}=P_{_{\textrm{IRS}}},
\end{align}\normalsize

\noindent\textbf{Diagonal loading for $\mathcal{P}^{(4)}_{_2}$:} 
Assume $\check{\lambda}$ is the maximum eigenvalue of $\mathcal{E}$. Rewrite $\mathcal{P}^{(4)}_{_2}$ using the diagonal loading $\check{\mathcal{E}}=\check{\lambda}\mbI-\mathcal{E}$ as\par\noindent\small
\begin{align}\label{St19}
\mathcal{P}^{(5)}_{_2}:~\underset{\mbb_{j}}{\textrm{maximize}}&~
\begin{bsmallmatrix} \bar{\mbb}_{j}\\1\end{bsmallmatrix}^{\mathrm{H}}
\underbrace{\begin{bsmallmatrix} (\check{\lambda}-\lambda^{\prime})\mbI+\check{\bOmega}^{(1)}(\mbb_{i}) &\uptau\bar{\mbb}_{i} \nonumber\\ \uptau \bar{\mbb}_{i}^{\mathrm{H}} & \check{\lambda}-2\uptau P_{_{\textrm{IRS}}} \end{bsmallmatrix}}_{\check{\mathcal{E}}(\mbb_{i})}
\begin{bsmallmatrix} \bar{\mbb}_{j}\\1\end{bsmallmatrix},\\
\textrm{subject to} & \quad \left\|\mbb_{1}\right\|^{2}_{2}=\left\|\mbb_{2}\right\|^{2}_{2}=P_{_{\textrm{IRS}}}.
\end{align}\normalsize
Define $\bar{\bar{\mbb}}_{j}= [\mbb^{\T}_{j} ~1~1]^{\T} = [\bar{\mbb}^{\T}_{j} ~1]^{\T} $, and $\bar{\mbI}=\left[\mbI_{L} ~\vdots ~\mathbf{0}_{L\times 2}\right]\in\realR{L\times (L+2)}$. The optimal beamforming gain $\mbb_{j}$ is the eigenvector corresponding to the dominant eigenvalue of $\check{\mathcal{E}}\left(\mbb_{i}\right)$, evaluated using the power iteration \cite{van1996matrix} at each iteration $t$ as follows:\par\noindent\small
\begin{equation}
\label{St20} 
\mbb^{(t+1)}_{j}=\sqrt{P_{_{\textrm{IRS}}}}~\bar{\mbI}\frac{\boldsymbol{\check{\mathcal{E}}}\left(\mbb^{(t)}_{i}\right)\bar{\bar{\mbb}}^{(t)}_j}{\left\|\boldsymbol{\check{\mathcal{E}}}\left(\mbb^{(t)}_{i}\right)\bar{\bar{\mbb}}^{(t)}_j\right\|_{2}},\; t \geq 0,\quad i\neq j \in \left\{1,2\right\}.
\end{equation}\normalsize
In each iteration, $\sqrt{P_{_{\textrm{IRS}}}}$ is considered to satisfy the norm-$2$
constraint, i.e., $\left\|\mbb\right\|^2_2=P_{_{\textrm{IRS}}}$. The projection 
is multiplied by $\bar{\mbI}$ to select only the first $L$ elements of the resulting vector.

To obtain quantized phase-shifts, we use the same bi-quadratic transformation process as \eqref{St16}-\eqref{St20}. Define new variables $\bar{\mbu}_{1}= [\mbu^{\T}_{1} ~1]^{\T}$ and $\bar{\mbu}_{2}=[\mbu^{\T}_{2} ~1]^{\T}$. The UQ$^2$P proposed as $\mathcal{P}^{(2)}_{_2}$ becomes
\par\noindent\small
\begin{align}\label{St22}
\mathcal{P}^{(6)}_{_2}:~\underset{\mbu_{j}\in\Omega^{L}_{M}}{\textrm{maximize}}&~
\begin{bsmallmatrix} \bar{\mbu}_{j}\\1\end{bsmallmatrix}^{\mathrm{H}}
\underbrace{\begin{bsmallmatrix} (\breve{\lambda}-\uplambda)\mbI+\check{\bOmega}^{(2)}(\mbu_{i}) &\breve{\uptau}\bar{\mbu}_{i} \\ \breve{\uptau} \bar{\mbu}_{i}^{\mathrm{H}} & \breve{\lambda}-2\breve{\uptau} L \end{bsmallmatrix}}_{\boldsymbol{\mathcal{K}}(\mbu_{i})}
\begin{bsmallmatrix} \bar{\mbu}_{j}\\1\end{bsmallmatrix},
\end{align}\normalsize
where $\breve{\lambda}$ is the diagonal loading parameter satisfying $\breve{\lambda}\mbI\succeq\check{\bOmega}^{(2)}$, $\uplambda$ is the maximum eigenvalue of $\check{\bOmega}^{(2)}$, and $\breve{\uptau}$ is the Lagrangian multiplier.

The constraint $\mbu_j\in\Omega^{L}_{M}$ (discrete phase-shift constellation) makes the problem $\mathcal{P}^{(6)}_{2}$  NP-hard 
\cite{kerahroodi2017coordinate,eamaz2023marli}. Therefore, an exhaustive search is required to identify good local quantized phase-shifts \cite{soltanalian2012computational}. We tackle this problem using MaRLI algorithm \cite{eamaz2023marli}, 
deploying the \emph{relaxation operator} in conjunction with the power iteration algorithm to approximate the solution.

Define $\bar{\bar{\mbu}}_{j}= [\mbu^{\T}_{j} ~1~1]^{\T} = [\bar{\mbu}^{\T}_{j} ~1]^{\T}$, the desired quantized IRS phase shifts $\mbu_{j}$ is therefore given at each iteration as\par\noindent\footnotesize
\begin{align}
\label{St23}
\mbu^{(t+1)}_{j}=\bar{\mbI}\left|\Tilde{\mbu}^{(t)}_{ij}\right|^{e^{-\nu_{1} t}} e^{\textrm{j} \frac{2 \pi}{M}\left(\left[\frac{M  \operatorname{arg}\left(\Tilde{\mbu}^{(t)}_{ij}\right)}{2 \pi}\right]+\left\{\frac{M  \operatorname{arg}\left(\Tilde{\mbu}^{(t)}_{ij}\right)}{2 \pi}\right\} e^{-\nu_{2} t}\right)},
\end{align}\normalsize
where $\nu_{1}$ and $\nu_{2}$ are parameters of relaxation operator and $\Tilde{\mbu}^{(t)}_{ij}=\boldsymbol{\mathcal{K}}\left(\mbu^{(t)}_{i}\right)\bar{\bar{\mbu}}^{(t)}_{j}$. Even though the final result is nearly identical to the quantized solution, we reapply quantization to ensure that the phases are accurately quantized. 
For continuous-valued (unquantized)  phase-shift scenario, i.e., $M\rightarrow\infty$, and $\left\{\nu_1, \nu_2\right\}=0$, the projection \eqref{St23} is 
$\mbu^{(t+1)}_{j}=\bar{\mbI}e^{\textrm{j} \operatorname{arg}\left(\boldsymbol{\mathcal{K}}\left(\mbu^{(t)}_{i}\right)\bar{\bar{\mbu}}^{(t)}_{j}\right)}$.

\noindent\textbf{To tackle $\mathcal{P}^{(1)}_{3}$ with respect to $\Tilde{\mbp}$:} We apply the same bi-quadratic transformation process as in \eqref{St16}-\eqref{St20} to obtain 
\par\noindent\footnotesize
\begin{align}\label{St25}
\mathcal{P}^{(3)}_{_3}:~\underset{\mbb_{j}}{\textrm{maximize}}&~
\begin{bsmallmatrix} \Tilde{\mbp}_{j}\\1\end{bsmallmatrix}^{\mathrm{H}}
\underbrace{\begin{bsmallmatrix} \lambda\mbI-\breve{\bOmega}(\Tilde{\mbp}_{i}) &\check{\uptau}\Tilde{\mbp}_{i} \\ \check{\uptau} \Tilde{\mbp}_{i}^{\mathrm{H}} & \lambda-2\check{\uptau} P_{_T} \end{bsmallmatrix}}_{\mathcal{B}(\Tilde{\mbp}_{i})}
\begin{bsmallmatrix} \Tilde{\mbp}_{j}\\1\end{bsmallmatrix},\nonumber\\
\textrm{subject to} & \quad \left\|\Tilde{\mbp}_{1}\right\|^{2}_{2}=\left\|\Tilde{\mbp}_{2}\right\|^{2}_{2}=P_{_T},
\end{align}\normalsize
where $\lambda$ is the diagonal loading parameter satisfying $\lambda\mbI\succeq\breve{\bOmega}$, and $\check{\uptau}$ is the Lagrangian multiplier. Define $\bar{\Tilde{\mbp}}_{j}= [\Tilde{\mbp}^{\T}_{j} ~1]^{\T} = [\bar{\mbu}^{\T}_{j} ~1]^{\T}$,  $\tilde{\mbI}=\left[ \mbI_{L}  ~\vdots ~\mathbf{0}_{L\times 1}\right] \in\realR{L\times (L+1)}$. 
The desired precoder $\Tilde{\mbp}=\operatorname{vec}\left(\mbP\right)$ is obtained at each iteration as\par\noindent\footnotesize	
\begin{equation}
\label{St26} 
\Tilde{\mbp}^{(t+1)}_{j}=\sqrt{P_{_{T}}}~\tilde{\mbI}\frac{\boldsymbol{\mathcal{B}}\left(\Tilde{\mbp}^{(t)}_{i}\right)\bar{\Tilde{\mbp}}^{(t)}_j}{\left\|\boldsymbol{\mathcal{B}}\left(\Tilde{\mbp}^{(t)}_{i}\right)\bar{\Tilde{\mbp}}^{(t)}_j\right\|_{2}},\; t \geq 0, \quad i\neq j \in \left\{1,2\right\}.
\end{equation}\normalsize
\textbf{Design Algorithm:} We iterate over \eqref{St20}, \eqref{St23}, and \eqref{St26} until the convergence  criteria  $|\textrm{SNR}_{_T}^{(t)}-\textrm{SNR}_{_T}^{(t-1)}| \leq \epsilon$, is met.
\vspace{-8pt}
\section{Numerical Experiments}
\vspace{-8pt}
\begin{figure}[t]
\centering
	\includegraphics[width=0.85\columnwidth]{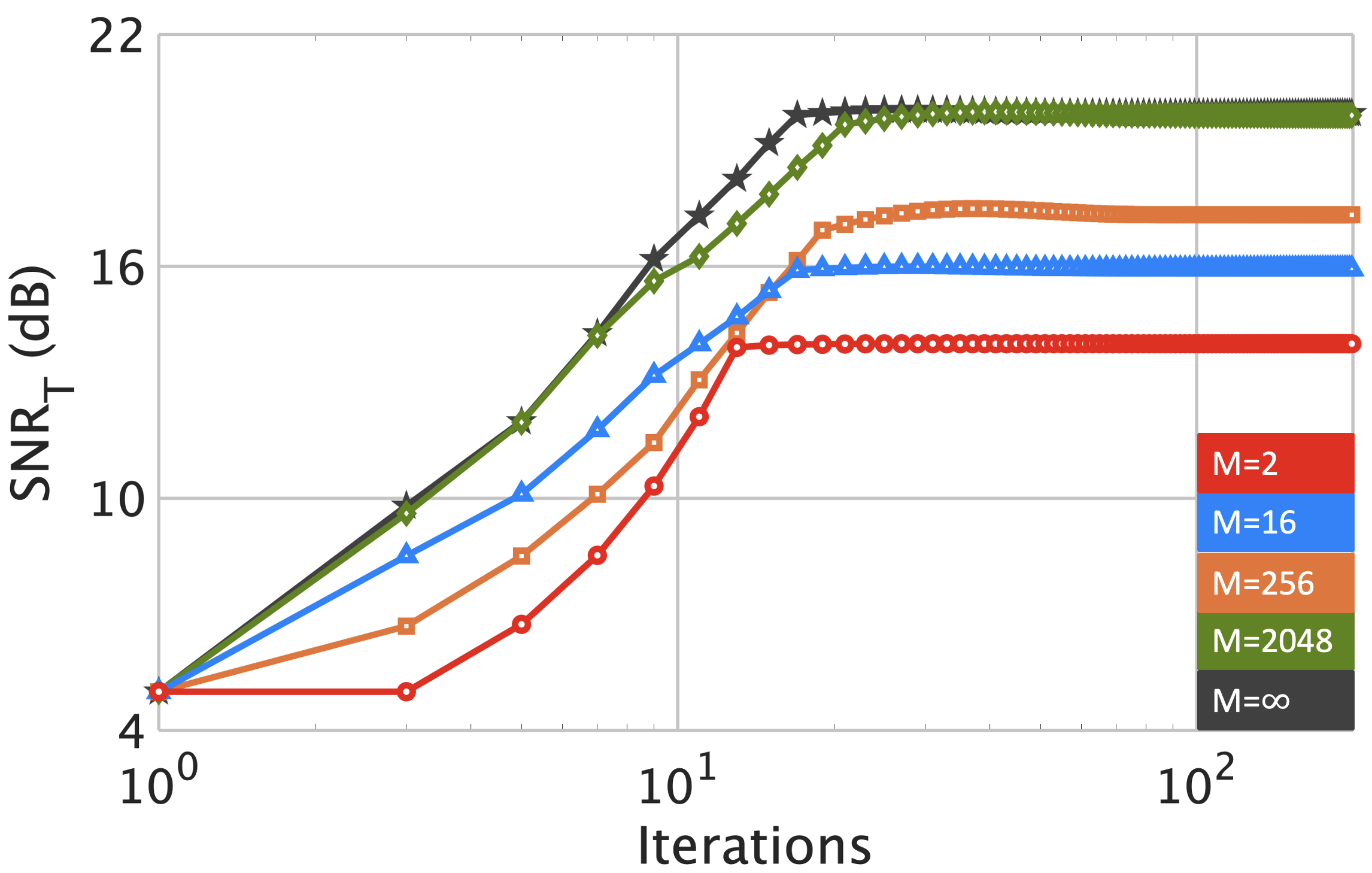}
	\caption{\footnotesize{The optimized   SNR$_{_T}$(dB) achieved for  $\beta=0.5$ , $P_{_T}=50$ dBm and $P_{_{\textrm{IRS}}}=30$ dBm versus $10^3$ iterations by jointly   designing an active IRS and DFBS precoder matrix. 
\vspace{-16pt} }}
 \label{fig_res}
\end{figure}
We validated our model and methods through numerical experiments. Throughout all simulations, we set the number of  DFBS antennas to $N=4$. The IRS had $L=16$  reflecting elements and the number of communications users was $K=5$. The radar target was located at  DoA $(45^{\degree},45^{\degree})$  and  $r_t=2500$m with respect to the IRS. The   CSI matrices are  generated according to the Rician  fading channel model~\cite{li2023minorization}.
The noise variances at the receivers of all communications users and DFBS were $\sigma_c^2=0$dBm. Following \cite{eamaz2023marli}, the MaRLI relaxation parameters were set to $\nu_{1}=1.2$ and $\nu_{2}=10^{-9}$. Fig.~\ref{fig_res} shows the achievable $\textrm{SNR}_{_T}$ with weight factor  $\beta=0.5$ through the iterations. At our algorithm, the  convergence threshold  was set to $\epsilon=10^{-3}$. The proposed  algorithm  converges to a higher value of SNR$_{_T}$ when we increase the number of  quantization bits $M$ because the size of the  feasible set in $\mathcal{P}_1$ grows larger. While MaRLI typically produces high-quality designs over discrete constellations, it is not a monotonic local optimizer \cite{eamaz2023marli}. The non-monotonic behavior of the SNR curves is presumably  attributed to this characteristic of the MaRLI algorithm.
\vspace{-8pt}
\section{Summary}
\vspace{-8pt}
We considered an IRS-ISAC setup and optimized the performance of both radar and communications receivers through the recently proposed tools in UQP optimization. In particular, our formulation includes the practical setting of quantization in IRS phase-shifts. Numerical experiments with both binarized and densely quantized levels indicate a convergence of our algorithm. 
\newcommand{\BIBdecl}{\setlength{\itemsep}{0.001 em}}
\bibliographystyle{IEEEtran}
\footnotesize{\bibliography{refs}}

\end{document}